\documentclass[prl,aps,twocolumn,superscriptaddress,showkeys,floatfix]{revtex4}

\usepackage{graphicx}
\usepackage[latin1]{inputenc}
\usepackage{hyperref}

\begin{document}

\title{Magnetic states at the Oxygen surfaces of ZnO and Co-doped ZnO}

\author{N. Sanchez, S. Gallego and M.C. Mu\~{n}oz}
\email{mcarmen@icmm.csic.es}
\affiliation{Instituto de Ciencia de Materiales de Madrid, CSIC, Madrid 28049, Spain}

\begin{abstract}
\textit{}
First principles calculations of the O surfaces  of Co-ZnO show that 
substitutional Co ions develop large magnetic moments 
which long-range coupling depends on their mutual distance.
The local spin polarization induced at the O atoms is
three times larger at the surface than in the bulk,
and the surface stability is considerably reinforced by Co.
Moreover, a robust ferromagnetic state is predicted at the Oxygen (0001) surface even in the 
absence of magnetic atoms.
The occurrence of surface magnetic moments  correlates with the number of
{\it p}-holes in the valence band of the oxide, and
the distribution of the magnetic charge is, even in the absence of spin-orbit 
interaction, highly anisotropic.
\end{abstract}

\date{\today}


\maketitle 
The prediction of room-temperature (RT) ferromagnetism (FM) in dilute magnetic oxides
generated a large interest, and at present there is wide experimental evidence 
for insulators such as ZnO, TiO$_2$ and SnO$_2$
when doped with just a few percent of magnetic transition-metal (TM) ions \cite{dms}.  
One of the most promising and extensively studied systems is Co-ZnO \cite{Ozgur05}. 
However, its magnetic properties and in particular RT-FM remain controversial, since 
they are not only strongly dependent on the preparation method, but also on the 
growth conditions of the samples \cite{coznoexp}.
The initial agreement between experimental observation of FM and calculations has 
evolved to the questioning of the possibility of RT-FM either mediated by valence-band
holes, or due to the percolation of magnetic polarons \cite{znoth,znoexp}.
Moreover, controversial issues need to be explained: the
independence of the Curie temperature (T$_C$) on the concentration of Co,
that magnetic moment (mm) per cation may exceed the limiting 
value established by Hund's rules or the time decay of the magnetization 
\cite{znorare}.
Nevertheless FM, though dependent on the growth conditions, has been reported in 
low-dimensional ZnO structures, even when undoped or doped with non-magnetic elements
\cite{NP}.

Despite ferromagnetic (F) order is mainly observed in low-dimensional structures 
with multiple surfaces and interfaces, 
to our knowledge there is not a theoretical study exploring how the surface affects 
the magnetic state of dilute oxides. 
In this Letter we investigate the local magnetic order at the polar
(0001) oriented surfaces of wurtzite ZnO. We show that Oxygens at 
Co-ZnO($000\overline{1}$) acquire large mm. The presence of the surface
enhances the spin polarization induced by Co atoms and  promotes surface 
magnetism. Furthermore, even in the absence of magnetic ions, a robust 
F state is predicted for the O-terminated (0001) ZnO 
surface. 

Our study of the spin resolved electronic structure is based on the  
\textit{ab-initio} pseudopotential density functional theory within the local
spin density approximation. We use the SIESTA package \cite{siesta} 
with basis sets formed by multiple-zeta polarized
localized numerical atomic orbitals (AO). In the case of Zn, we choose 
double-zeta (DZ) basis for the $s$ and $d$ AOs plus a single-zeta (SZ) $p$ AO.
The same type of basis set is used for Co, whereas for O we employ DZ 
$s$ and $p$ AOs plus a SZ $d$ AO. More details about the conditions of the
calculations can be found elsewhere \cite{prb-zro2}.

\begin{figure}[htbp]
\centerline{\includegraphics[width=0.5\textwidth,angle=-90,clip,viewport=0 80 570 791]{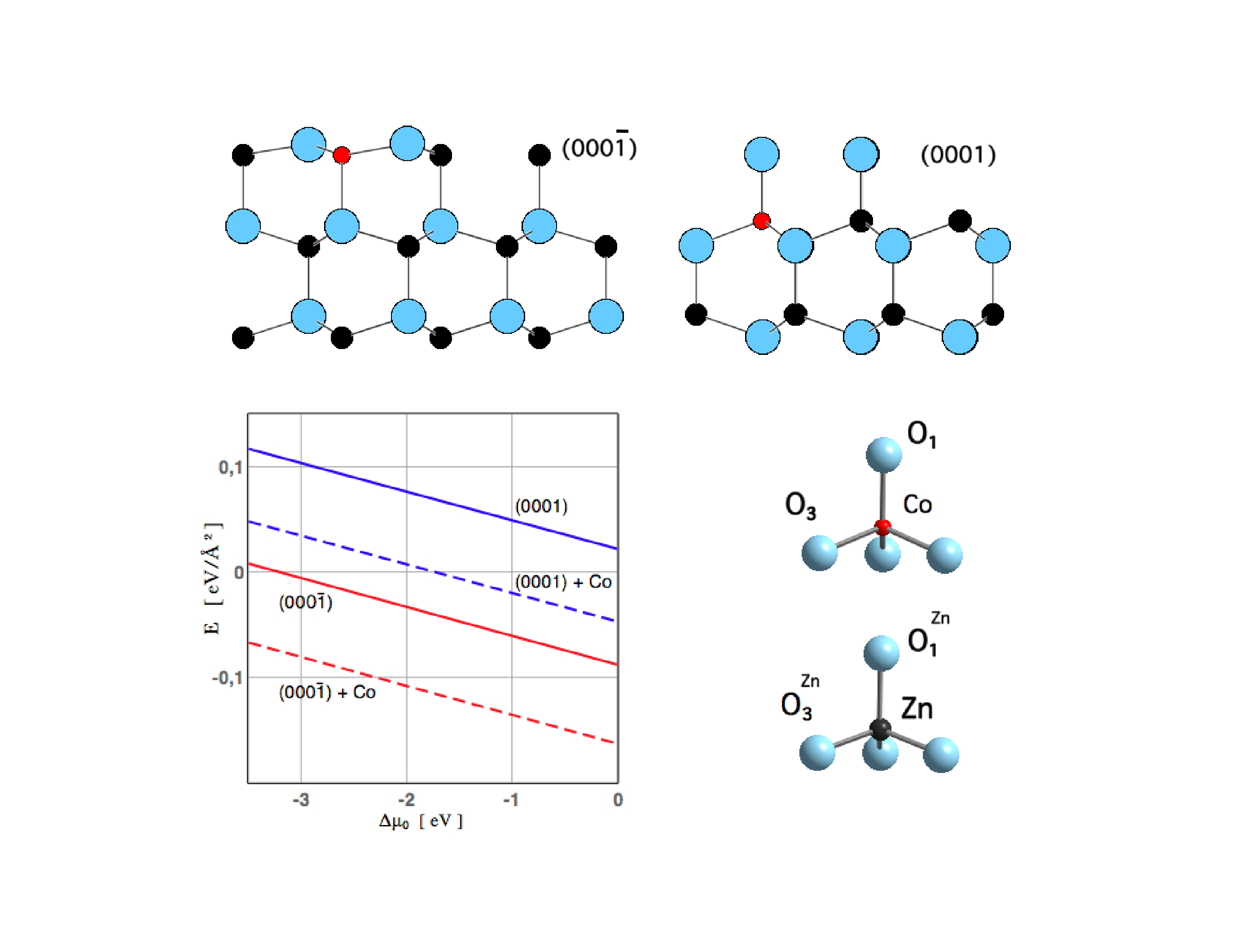}}
\caption{\label{struct}(color) (Top) Side views of all possible ($000\overline{1}$) and (0001)
wurtzite ZnO surfaces, indicating the position of substitutional Co
at the Co-doped O terminations modelled here. Our labels for the atomic
sites are shown in the coordination units at the bottom right side.
(Bottom) Relative surface free energy as a function of the O partial pressure for the
undoped (solid lines) and Co-doped (dashed) ($000\overline{1}$) and (0001) surfaces. 
For negative energy values the O-termination is more stable than the Zn- one.}
\end{figure}

Since the wurtzite structure is not centrosymmetric, thin layers grown along 
the crystallographic {\it c}-axis may present either the [0001] or the 
[$000\overline{1}$] growth direction depending on the substrate orientation.
Along this direction, pure hexagonal O and Zn layers alternate, so that the
resulting surfaces can be either Zn- or O-terminated, as shown in
figure~\ref{struct}. We model them by
periodically repeated slabs containing between 15 and 17 ZnO planes, and separated by a 
vacuum region of at least 15 $\AA$. 
Bulk-like behaviour is always attained at the innermost central layers.
We use a ($1 \times 1$) two-dimensional (2D) unit-cell to study the 
undoped bulk and surfaces, and a ($2 \times 2$) 2D-cell containing four atoms per plane for the 
Co doped structures. Substitution of one and two Zn atoms by Co gives 3.6 and 7.5\% of 
dopant concentration, respectively. We calculate both symmetric and asymmetric 
slabs about the central plane. For the former, the two 
sides of the slab correspond respectively to the
inequivalent (0001) and ($000\overline{1}$) surfaces,
which are either both Zn-ended or O-ended;
for the second, a Zn and an O surface layer are at opposite ends of the slab. 
For asymmetric slabs dipolar corrections are applied. 
The atomic positions are allowed to relax until the forces on the atoms are less than 0.05 eV/$\AA$,
and usually below 0.03 eV/$\AA$. Brillouin Zone
(BZ) integrations have been performed on a $4 \times 4 \times 1$ Monkhorst-Packard supercell
that includes as much as 1800 k-points in the 1/4 irreducible 2D BZ, although
convergence has been verified using $8 \times 8 \times 1$ cells.

The stability of a given surface depends on the partial Oxygen pressure.
For a surface in thermodynamic equilibrium, the surface free energy 
can be obtained from an {\it ab-initio} total energy calculation in 
combination with a thermodynamic formalism \cite{ssrep}. In Fig~\ref{struct} the calculated 
relative stability of the different defect-free surfaces with [0001] 
orientation as a function of the Oxygen chemical potential is shown. The O-terminated 
($000\overline{1}$) surface is found to be stable over the entire admissible range of 
chemical potentials, in good agreement with experiments \cite{Tsuka05}.
Contrary, for the (0001) surface the most stable surface
corresponds to the Zn-termination. However, along the [0001]
direction, rough surfaces associated to Zn vacancies and incomplete Oxygen terminations
have been observed, and even smooth (0001) O-terminated surfaces are formed under optimum 
growth conditions \cite{znosurf}. In the following, we will concentrate in 
the O- surfaces. 

\begin{table}[htbp]
\centering
\caption{Magnetic moment (mm, in $\mu_B$) and charge difference with respect to the bulk 
($\Delta$Q$=Q_b - Q$, where $Q_b$ refers to bulk Co or ZnO) of Co and its O neighbours for
bulk Co-ZnO, and at the Co and two different O sites of the three outermost 
layers in the O-ended ($000\overline{1}$) and (0001)
surfaces, the last either unrelaxed (U) or relaxed (R). 
The O surfaces for undoped ZnO are also shown. 
Layers are numbered from the surface (L1) to the bulk.
}
\label{tmagmo}
\renewcommand{\arraystretch}{1.2} 
\renewcommand{\tabcolsep}{0.3pc} 
\begin{tabular}{lp{6mm}l|rrp{1mm}rp{1mm}rr}
   &    &    &   O$_{L1}$  & O$^{Zn}_{L1}$ && \multicolumn{1}{c}{Co$_{L2}$}  && O$_{L3}$  & O$^{Zn}_{L3}$ \\
\hline
mm & Co- & Bulk &             0.11 &      && 2.71 && 0.12 &     \\
&&($000\overline{1}$)&        0.45 & 0.03 && 2.90 && 0.22 & 0.03 \\
 & & (0001)$_U$ &             1.39 & 1.39 && 2.84 && 0.20 & 0.04 \\
 & & (0001)$_R$ &             0.80 & 1.33 && 2.05 && 0.17 & 0.03 \\[1mm]
&Und.&($000\overline{1}$)&         & 0.00 &&      &&      & 0.00 \\
 & & (0001) &                      & 1.40 &&      &&      & 0.08 \\
\hline
$\Delta$Q & Co- & Bulk &     -0.03 &       && -0.80 && -0.03 &      \\
&&($000\overline{1}$)&       -0.18 & -0.06 && -0.83 && -0.04 &  0.01 \\
 & & (0001)$_U$ &            -0.58 & -0.58 && -0.04 &&  0.04 &  0.02 \\
 & & (0001)$_R$ &            -0.69 & -0.57 && -0.65 &&  0.09 &  0.01 \\[1mm]
&Und.&($000\overline{1}$)&         &  0.11 &&       &&       &  0.01 \\
 & & (0001) &                      & -0.58 &&       &&       & -0.03 \\
\end{tabular}
\end{table}

\begin{figure}[htbp]
\centerline{\includegraphics[width=0.5\textwidth]{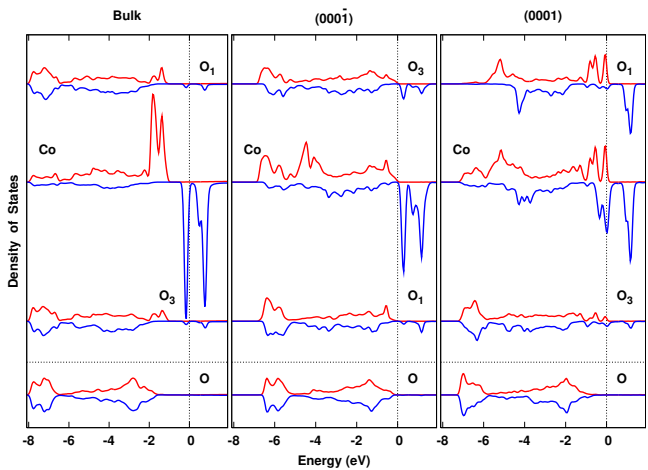}}
\caption{\label{dosp2} LDOS for Co and the O atoms bonded to Co at
bulk Co-ZnO (left), and at the three uppermost layers of the O-ended
$(000\overline{1})$ (middle) and (0001) (right) surfaces of 3.6\% Co-doped ZnO.
The top curves refer to the surface plane (L1 in table \protect\ref{tmagmo}),
while the bottom ones show the bulk-like LDOS of O at the central layers
of each slab.
Positive (negative) values correspond to majority (minority) spin states. 
Energies are referred to E$_F$.}
\end{figure}

First, we analyze the Co-doped O-terminated ($000\overline{1}$) surface.
For a 3.6\% Co concentration, Co substitutes a Zn in the cation subsurface layer
(see figure~\ref{struct}).
The surface O atoms have an unsaturated dangling bond and are bonded
either only to Zn (O$^{Zn}_3$ atoms), or to one Co and two Zn atoms of the subsurface 
layer (O$_3$ atoms).
The substitution of Zn by Co does not involve a relevant variation
of bond distances. 
However, besides the spin-polarization, Co-doping significantly reduces the surface energy,
increasing the surface stability as shown in figure~\ref{struct}.
Fig.~\ref{dosp2} provides the layer projected density of states (LDOS) 
of the Co and Co-bonded O atoms at the three uppermost layers,
together with that of an inner bulk-like O.
For comparison, the corresponding LDOS for bulk Co-ZnO are shown on the left.
The calculated mm and Mulliken charges are given in table~\ref{tmagmo}.
In both, surface and bulk, we find a magnetic ground state which remains 
insulating. In addition, Co shows a spin-polarized LDOS with 
a large mm close to
3$\mu_B$, the maximum value allowed by Hund's rule. However, there are
significant differences when Co is located near the surface or in the bulk.
The majority-spin Co 3$d$ states are fully occupied 
in both cases although the hybridization with the O 2$p$ valence band
is stronger for the subsurface Co, which does not show the well 
defined $e$ and $t$2 levels seen in bulk Co. 
Also, while the highest occupied levels in bulk Co
are the minority-spin $e$ states,
both $e$ and $t$2 minority states are empty in the subsurface Co. 
Consequently, Co exchange splitting and Co-O hybridization are
much larger when Co is close to the surface, leading to a 
larger induced spin-polarization of the surface Oxygens. 
In addition, the table suggests certain correlation between mm and charge transfer:
surface Oxygen atoms bonded to Co with large mm show a significant decrease 
of the ionic charge.

There is a gradual evolution of the above described properties for subsurface and
bulk Co as the Co distance to the surface increases.
Moreover, although Co induced changes are mostly localized around the impurity,
their effect extends more at the surface than in the bulk. For example, the 
mm of the O atoms closer to Co but with no direct bond to it is
0.03 $\mu_B$ when Co is in the subsurface layer, while it drops to zero
when Co is in the bulk, resulting in a total mm per unit cell of  4.76 (3.21) 
$\mu_B$ at the surface (bulk). 
This unambigously demonstrates that the breaking symmetry of the surface 
promotes magnetism.

We have studied the difference between F and antiferromagnetic (AF) ordering substituting an additional 
Zn by Co at different cation layers (7.5\% of Co concentration).
The local main features depicted in Fig.~\ref{dosp2} are not altered.
AF coupling between the Co impurities is favoured whenever Co atoms
are separated by more than a ZnO unit, while F coupling becomes stable if there 
is only an O between them. However, 
even for an AF alignment of Co atoms,
the existence of uncompensated O mm at the surface leads to a net F structure.
Nevertheless, the energy difference between states with F and AF Co coupling
is always too small, $\sim 10$ meV, 
to account for the experimental T$_C$.
Addition of codoping does not modify
substantially the magnitude of the coupling, in agreement with previous 
calculations\cite{spaldin}.

Regarding the 3.6\% Co-doped (0001) O-ended surface,
again the presence of Co atoms close to the surface remarkably increases
its stability, in agreement with recent experiments \cite{stability},
leading to a range of the chemical potential for which the O-termination
is the most stable (see figure \ref{struct}). 
Oxygen atoms have three unsaturated dangling bonds, and therefore they are onefold 
coordinated only to the metal atom underneath, either a Co (O$_1$ atoms) or a Zn (O$^{Zn}_1$ atoms). 
Due to the low coordination, which is clearly evidenced in the large
charge loss, important relaxations occur. For this 
reason, the unrelaxed and relaxed values are provided in table~\ref{tmagmo}.
The main effect of relaxation is to modify the Co-O bond distances, which can become 
as small as 1.5$\AA$ for the O$_1$ atoms. The short bond length modifies 
largely the Co mm. In fact, the Co exchange splitting becomes smaller than that for an 
impurity in the bulk,
as shown in the rightmost panel of figure \ref{dosp2}. 
The Fermi level (E$_F$) crosses the valence band, leading to a metallic surface.
On the other hand, the induced mm of surface Oxygen is almost twice than at
the ($000\overline{1}$) surface, reflecting the stronger hybridization to Co.
Nevertheless, the more stricking effect is that
all the surface Oxygens develop a mm 
independent of the magnetic character of the cation neighbour, 
which is even larger for the O$^{Zn}_1$ atoms, 1.3$\mu_B$. 
Further, the LDOS of the two inequivalent surface O (not shown here) show 
different shapes, pointing to a dissimilar origin of the mm.
In summary, it seems like there are two different types of mm for O: one induced
by hybridization with the impurity $d$ states, and other which can not be associated 
to the $d$ electrons. 

\begin{figure}[htbp]
\centerline{\includegraphics[width=0.5\textwidth]{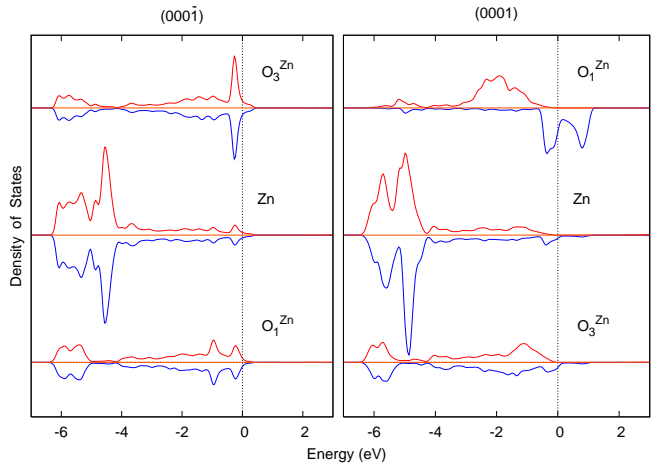}}
\caption{\label{dosp14}Same as figure~\ref{dosp2} for
the three topmost layers of the O-ended undoped 
ZnO ($000\overline{1}$) and (0001) surfaces.}
\end{figure}

\begin{figure}[htbp]
\centerline{\includegraphics[width=0.5\textwidth,clip,viewport=175 356 470 510]{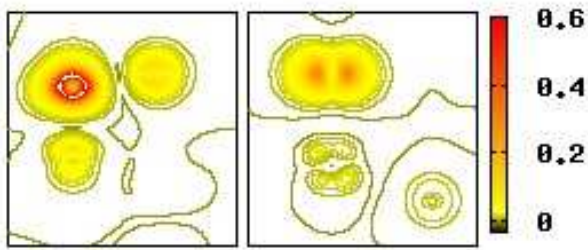}}
\caption{\label{fchdens}(color) SDD at the two topmost surface planes of the O-ended 
(left) Co-ZnO($000\overline{1}$) surface around the Co site and
(right) undoped ZnO(0001) surface.}
\end{figure}

To get a further insight on this idea, 
we have calculated the ($000\overline{1}$) and (0001) O surfaces
of undoped ZnO allowing for the spin degree of freedom. 
We find that the configuration with 
ferromagnetically ordered spin moments is favoured at the (0001) surface.
The energy reduction due to the spin polarization is 0.6 eV per O, a value 
in the range of those obtained for magnetic bulk oxides. 
The LDOS of the undoped surfaces are displayed in 
figure~\ref{dosp14}, and the mm and ionic charge differences
are also included in table~\ref{tmagmo}. 
Oxygen at the ($000\overline{1}$) surface presents an occupied large
peak close to E$_F$ but it has not an appreciable  
spin-polarization. On the other hand, a large splitting is observed in
the Oxygen atoms at the (0001) surface, which develop a mm
very similar to that of O bonded to Zn in the Co-doped surface.
Remarkably, the bands crossing E$_F$ correspond to minority-spin levels,
leading to a half-metallic system
with charge compensating holes of well defined spin polarization.
The lower O coordination of the (0001) surface as compared to the
($000\overline{1}$) one leads to a lower charge transfer 
and then to a higher number of 2$p$ holes in the O valence band. 
This different response proves that it is 
necessary a critical number of 2$p$ holes for achieving a magnetic surface.
This result corroborates previous findings of correlation between ionic 
charge and mm on a systematic study of magnetic surfaces of highly ionic oxides\cite{jphysc}.
Another peculiarity of Oxygen mm induced by creating 2$p$ holes is the
highly anisotropic distribution of the magnetic charge, which involves
only $p$-orbitals along specific directions. 
The crystal field determines where holes are localised: while the majority
spin is completely filled and thus adopts the spherical symmetry of the bulk,
there is an anisotropic distribution of the minority charge, and holes
mainly reside in the $p_{xy}$ orbitals of the (0001) O surface.
This is seen in figure~\ref{fchdens}, which depicts the spin charge density 
differences (SDD, total charge density minus the superposition of atomic charge
densities) for the Co-doped $(000\overline{1})$  and the undoped (0001)
surfaces, evidencing the different orbital
contributions to the mm when induced by the 2$p$ holes or
by Co-O hybridization.

In conclusion, the proximity of a surface to a substitutional Co in ZnO enhances 
the magnetization induced by the magnetic impurity and even for an AF alignment of 
Co impurities the surface may show uncompensated spins ferromagnetically ordered,
although this effect alone does not justify a high T$_C$. 
The presence of Co atoms close to the surface remarkably
increases its stability.  In addition, in the absence of magnetic atoms,
$p$-magnetic states can still develop whenever a critical value of $p$-holes
is exceeded, with a highly anisotropic distribution of the magnetic charge. 
The existence of oxygen polarised large surface areas can account
for some of the spontaneous magnetization reports observed in several ZnO structures,
even in the absence of magnetic doping. Also, it can explain why
the magnetic state is intimately connected to the actual structure of the sample
and the growth conditions.
Oxygen $p$-magnetism associated to uncompensated charge in ionic oxides
seems to be a general phenomenon which can be related to early reports about the fact
that cation-deficient CaO or SrO should be 
half-metallic ferromagnets \cite{cao}, the recent proposal of magnetizing oxides
by substituting nitrogen for oxygen \cite{Elfimov07}, as well as to the magnetic
properties of cation vacancies in II-VI semiconductors \cite{catvac}.
Recently, magnetic states have been reported 
at the interface between two non-magnetic oxides \cite{magioxid}. This
opens new perspectives for the development of the new fascinating field
of oxides heterostructures in which the violation of charge compensation 
may give rise to new magnetic or superconducting states \cite{intoxid}.

\begin{acknowledgments} This research was supported by 
the Spanish Ministry of Science and Technology under Project
MAT2006-05122.
\end{acknowledgments}


\begin{thebibliography}{28}
\bibitem{dms}
 J.M.D. Coey, {\it Curr. Opin. Solid State Mater. Sci.} {\bf 10}, 83 (2006)
 and references therein.
\bibitem{Ozgur05}
 \"U. Ozgur {\it et al.}, {\it J. Appl. Phys.} {\bf 98}, 41301 (2005).
\bibitem{coznoexp}
 S.A. Chambers, Surf. Sci. Rep. {\bf 61}, 345 (2006);
 J. Cui, and U. Gibson, {\it Phys. Rev. B} {\bf 74}, 045416 (2006);
 P. Sati {\it et al.}, {\it Phys. Rev. Lett.} {\bf 98}, 137204 (2007).
 Q. Xu {\it et al.}, {\it Appl. Phys. Lett.} {\bf 92}, 082508 (2008)
\bibitem{znoth}
 K. Sato, H. Katayama-Yoshida, {\it Semicond. Sci. Technol.} {\bf 17}, 367 (2002);
 L. Petit {\it et al.}, {\it Phys. Rev. B} {\bf 73}, 45107 (2006);
 S. Hu, S. Yan, M. Zhao and L. Mei, {\it Phys. Rev. B} {\bf 73}, 245205 (2006);
 C.H. Patterson, {\it Phys. Rev. B} {\bf 74}, 144432 (2006).
\bibitem{znoexp}
 C. Song {\it et al.}, {\it Phys. Rev. B} {\bf 73}, 24405 (2006);
 A.J. Behan {\it et al.}, {\it Phys. Rev. Lett.} {\bf 100}, 47206 (2008).
\bibitem{znorare}
 N.H. Hong {\it et al.}, {\it Phys. Rev. B} {\bf 72}, 45336 (2005);
 J.M.D. Coey, {\it J. Appl. Phys.} {\bf 97}, 10D313 (2005);
 A. Che Mofor {\it et al.}, {\it Appl- Phys. Lett.} {\bf 87}, 62501 (2005).
\bibitem{NP}
 A.S. Risbud {\it et al.}, {\it Phys. Rev. B} {\bf 68}, 205202 (2003);
 M.A. Garc\'{\i}a {\it et al.}, {\it NanoLetters} {\bf 7}, 1489 (2007).
\bibitem{siesta}
 P. Ordej\'on, E. Artacho and J.M. Soler, {\it Phys. Rev. B} {\bf 53}, R10441 (1996);
 J.M. Soler {\it et al.}, {\it J. Phys.: Condens. Matter} {\bf 14}, 2745 (2002).
\bibitem{prb-zro2}
 J.I. Beltr\'an {\it et al.}, {\it Phys. Rev. B} {\bf 68}, 075401 (2003).
\bibitem{ssrep}
 M.C. Mu\~noz {\it et al.}, {\it Surf. Sci. Rep.} {\bf 61}, 303 (2006).
\bibitem{Tsuka05}
 A. Tsukazaki {\it et al.}, {\it Nat. Mater.} {\bf 4}, 42 (2005).
\bibitem{znosurf}
 O. Dulub, L.A. Boatner, U. Diebold, {\it Surf. Sci.} {\bf 519}, 201 (2002);
 Y. Ding, Z.L. Wang, {\it Surf. Sci.} {\bf 601}, 425 (2007).
\bibitem{spaldin}
 P. Gopal and N.A. Spaldin, {\it Phys. Rev. B} {\bf 74}, 094418 (2006).
\bibitem{stability}
 H. Matsui, H. Tabata, {\it Phys. Rev. B} {\bf 75}, 14438 (2007); 
 C. Zeng {\it et al.}, {\it Phys. Rev. Lett.} {\bf 100}, 66101 (2008).
\bibitem{jphysc}
 S. Gallego, J.I. Beltr\'an, J. Cerd\'a, and M.C. Mu\~noz, 
 {\it J. Phys.: Condens. Matter} {\bf 17}, L451 (2005).
\bibitem{cao}
 W.E. Pickett, {\it Phys. Rev. Lett.} {\bf 77}, 3185 (1996);
 I.S. Elfimov, S. Yunoki, G.A. Sawatzky, {\it Phys. Rev. Lett.} {\bf 89}, 216403 (2002);
 R. Pentcheva, W.E. Pickett, {\it Phys. Rev. B} {\bf 74}, 035112 (2006);
 J. Osorio-Guill\'en {\it et al.}, {\it Phys. Rev. Lett.} {\bf 96}, 107203 (2006).
\bibitem{Elfimov07}
 I.S. Elfimov {\it et al.}, {\it Phys. Rev. Lett.} {\bf 98}, 137202 (2007).
\bibitem{catvac}
 T. Chanier {\it et al.}, {\it Phys. Rev. B} {\bf 73}, 134418 (2006);
 T. Dietl, {\it Phys. Rev. B} {\bf 77}, 085208 (2008).
\bibitem{magioxid}
 A. Brinkman {\it et al.}, {\it Nature materials} {\bf 6}, 493 (2007).
\bibitem{intoxid}
 A. Tsukazaki {\it et al.}, {\it Science} {\bf 315}, 1388 (2007);
 N. Reyren {\it et al.}, {\it Science} {\bf 317}, 1196 (2007);
 W. Eerenstein, N.D. Mathur, J.F. Scott, {\it Nature} (London) {\bf 442}, 759 (2006).  
\end{thebibliography}
\end{document}